# Monte-Carlo Simulations of Radiation-Induced Activation in a Fast-Neutron and Gamma-Based Cargo Inspection System

**B. Bromberger**[a*]**, D. Bar**[b]**, M. Brandis**[b]**, V. Dangendorf**[a]**, M. B. Goldberg**[b]**,
**F. Kaufmann**[a]**, I. Mor**[a]**, R. Nolte**[a]**, M. Schmiedel**[a]**, K. Tittelmeier**[b]**, D. Vartsky**[a] **and
**H. Wershofen**[a]

[a] *Physikalisch-Technische Bundesanstalt (PTB),*
   *38116 Braunschweig, Germany*
[b] *Soreq NRC,*
   *81800 Yavne, Israel*

   *E-mail*: benjamin.bromberger@ptb.de

ABSTRACT: An air cargo inspection system combining two nuclear reaction based techniques, namely Fast-Neutron Resonance Radiography and Dual-Discrete-Energy Gamma Radiography is currently being developed. This system is expected to allow detection of standard and improvised explosives as well as special nuclear materials.

An important aspect for the applicability of nuclear techniques in an airport inspection facility is the inventory and lifetimes of radioactive isotopes produced by the neutron and gamma radiation inside the cargo, as well as the dose delivered by these isotopes to people in contact with the cargo during and following the interrogation procedure. Using MCNPX and CINDER90 we have calculated the activation levels for several typical inspection scenarios. One example is the activation of various metal samples embedded in a cotton-filled container.

To validate the simulation results, a benchmark experiment was performed, in which metal samples were activated by fast-neutrons in a water-filled glass jar. The induced activity was determined by analyzing the gamma spectra.

Based on the calculated radioactive inventory in the container, the dose levels due to the induced gamma radiation were calculated at several distances from the container and in relevant time windows after the irradiation, in order to evaluate the radiation exposure of the cargo handling staff, air crew and passengers during flight. The possibility of remanent long-lived radioactive inventory after cargo is delivered to the client is also of concern and was evaluated.

KEYWORDS: MCNPX; CINDER90; Geant4; Simulation; Benchmark; Activation; Dose; Gamma; Fast Neutron; Cargo Inspection

---

[*] Corresponding author.

# Contents

## 1. Introduction

To meet emerging and future challenges posed by terrorist activities, a bi-national German-Israeli collaboration is working on a project entitled "Automatic Contraband-in-Cargo Interrogation System" (ACCIS), whose goal is the research and evaluation of a new type of cargo inspection system on a laboratory scale, as well as issues related to implementation of the technique, radiation safety, legal implications and public acceptance. The screening techniques employed are based on two independent physical methods, Dual Discrete-Energy γ-Radiography (DDEGR) [1] and Fast-Neutron Resonance Radiography (FNRR) [2], combined into a single facility. This combination is expected to enable detection of special nuclear materials, as well as standard and improvised explosives hidden in medium-sized containers or other sealed cargo.

As these methods call for a dual-line gamma spectrum on the one hand (DDEGR) and a broad energy fast-neutron spectrum on the other (FNRR), a suitable radiation source is required. This can be achieved by means of an ion accelerator producing ~5-7 MeV deuterons, which induce two discrete gamma lines at 4.43 MeV and 15.11 MeV, as well as a quasi-continuous energy neutron spectrum ranging from ~1 MeV to around 20 MeV [1], [3] via the $^{11}B(d,n+\gamma)^{12}C$ nuclear reaction .

Figure 1 shows an artists' view of a possible future deployable interrogation system. In this setup, a single radiation source with a suitable collimator creates three fan beams of neutrons and gamma rays which penetrate the container at different angles. This will permit performing few-view tomography on the scanned goods, to provide 3-dimensional detection capabilities [3].

The neutron fluence required for a single radiograph was estimated to be $10^6$ cm$^{-2}$ at the source side of the container surface, taking into account anticipated image statistics, the desired detection sensitivity, average cargo size and density, as well as issues related to neutron scattering and capture in such a container. For complete screening of a 2 m long container, around 20 single scan positions along the axis of motion (see Figure 1) will be required.

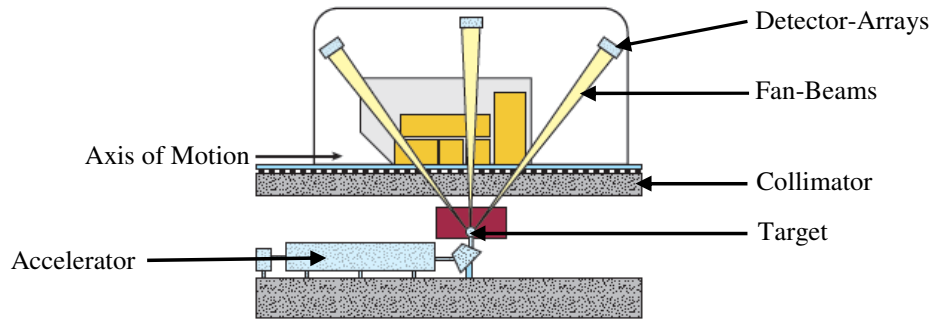

**Figure 1.** Schematic view of the proposed cargo scanner

At the given fluence, nuclear-reaction-induced activation by neutrons and gamma-rays on the cargo raise substantial questions concerning radiation safety, particularly with respect to expected doses to the cargo handlers and the possible need to let the induced radioactivity decay before loading the cargo onto a plane or truck. In this context, it should be noted that ~50 % of all commercial air cargo is shipped in passenger aircraft. Thus, one needs to ensure that the expected dose to passengers on the flight in question is negligible or at least in compliance with the prescribed legal limits. Furthermore, it must be assured that induced activation in the goods is negligible when they are delivered to the recipient. Another relevant question pertains to the possibility that the containers themselves accumulate long-lived radioactive isotope inventories during repeated inspections within their life cycle.

As no system is yet available for performing experimental measurements, these scenarios had to be simulated, in order to determine induced activation as well as dose to environment. For this purpose, an appropriate Monte-Carlo code had to be chosen. The requirements for the code of choice are a well validated neutron physics implementation, including access to established evaluated data files, a possibility of isotope production via neutron capture, inelastic neutron scattering, elastic scattering and other, more complex reaction channels, including multiple reactions. Moreover, the ability to calculate the neutron-induced activity by means of the produced isotopes and time-dependence of the radioactive inventory are relevant selection criteria for the code.

From this point of view, only the two most frequently employed codes, namely MCNPX [4] and Geant4 [12] were considered to be viable candidates. However, since Geant4 was originally written for high energy physics purposes, its neutron physics and radioactive decay features are not yet well validated. In addition, the built-in features concerning the monitoring of time-dependent decay processes are limited. For example, isotope production by neutron and gamma interactions is included, but the induced activation and its time dependence need to be calculated by hand, by means of the number of transmuted nuclei and their respective half-lives. MCNPX, on the other hand, was explicitly developed for neutron-based simulations and although isotope production is not included, additional burn-up codes are available, such as CINDER90 [5] or Monteburns [7], which offer the possibility of calculating automatically induced activities and their successive decay. These make use of MCNPX's average cell flux tally (F4), the given geometry and material definitions. Furthermore, MCNPX is an input-file based code which makes it much easier to set up a simulation, in contrast to Geant4, which is more like a software development kit. Therefore a simulation framework consisting of the Monte-Carlo codes MCNPX and CINDER90 was employed. Furthermore, in order to validate the simulation codes, their database and the simulation procedure, a benchmark experiment was performed.

## 2. Benchmark Experiment

Monte-Carlo simulations, even those using verified and long proven code systems are prone to errors in their results due to minor mistakes in their inputs and faulty handling. Therefore it is recommended to benchmark or validate a simulation either against another simulation with a different Monte-Carlo code or, preferably, against experimental data. As no applicable data could be found in the literature, it was decided to perform our own benchmark experiment, since an ion accelerator for the irradiation of samples and a high-resolution, low-background germanium detector for gamma spectrometry were both available.

To this end, samples of different elements were prepared and irradiated for around 30 minutes employing an 8 MeV neutron beam. Subsequently, the gamma activity of the samples was measured with a shielded, well-type germanium detector, the gamma spectrum being accumulated for around 30 minutes. In the final step, the entire benchmark experiment was simulated and the results compared to experiment.

### 2.1 Experimental Setup

The irradiation samples of aluminum, copper, iron and silver were machined in cylinder form, with dimensions of 30 mm in height and 9 mm in diameter. Another sample, consisting of manganese powder was prepared, filling cylindrical vials with inner dimensions of 6.1 mm in diameter and 33.3 mm in height, to fit into the well of the germanium detector.

The irradiation took place at the PTB Ion Accelerator Facility (PIAF) [6], employing the 19 MeV CV-28 cyclotron built by TCC (The Cyclotron Corporation) to accelerate deuterons to an energy of around 5 MeV, in order to produce ~8 MeV neutrons in a deuterium gas target, utilizing the $D(d,n)^3He$ nuclear reaction.
To characterize flux and energy distribution of the neutron beam, a neutron spectrometer consisting of a 10.16 cm diameter and 2.54 cm long cylindric NE-213 scintillator and a Photonis XP2020 photomultiplier tube was employed. The neutron fluence to the samples was referenced to a measurement with a $^{238}U$ fission ionisation chamber, which was placed at a distance of 6 cm from the center of the gas target. Using the $^{238}$U cross-section at the measured neutron energy as a standard, a NE213 fluence transfer monitor as well as published values for the D(d,n) angular dependence of the energy and neutron flux were used to determine the neutron fluence to the samples during the irradiation period.

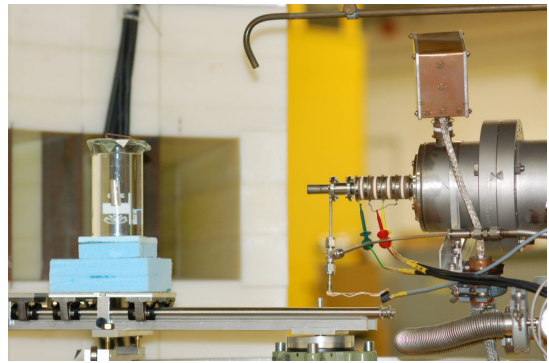

**Figure 2.** Photograph of the irradiation setup

Before irradiation, the samples were placed inside a water-filled glass beaker, to partly moderate the incoming fast neutrons, thereby making the experiment more sensitive to the transport features of the MCNPX code. They were positioned at a distance of 20 cm in front of the deuterium gas target. A photograph of this setup can be seen in Figure 2.

**2.2 Activation Measurements**

After 30 minutes of irradiation in the neutron beam, the samples were taken out of the cyclotron room and transported to the germanium detector (this step took about 5-6 minutes). The gamma detector was a high-precision, p-type germanium well detector with a relative efficiency of 65.3 %, consisting of a 480 cm$^3$ germanium crystal. The detector was shielded against environmental background radiation by a 20 cm thick layer of low $^{210}$Pb-containing lead and a 5 mm layer of electrolytic-copper. Furthermore, the well is flushed with nitrogen gas coming out of the Dewar, to remove radon and its daughter isotopes from the chamber. The cylindrical well has dimensions of 1 cm in diameter and 6.2 cm in length [8]. After the samples were positioned in the center of the detector well, the spectrum of the emitted gamma rays was measured employing the data acquisition software InterWinner [9]. Taking into consideration the correct coincidence-summing coefficients and self-attenuation corrections (calculated by employing the software package GeSpeCor [10]) this software calculates the activity of a specific isotope in the sample by means of the total counts in the corresponding gamma peaks, the half-life of the corresponding isotope, the time duration between end of irradiation and end of spectrum measurement, as well as the calibrated efficiency of the detector at the gamma energy in question.

**2.3 Simulation and Results**

For the simulation of this benchmark experiment, the geometry of the irradiation setup was implemented into MCNPX. To determine simulation parameters such as the neutron energy distribution and angular distribution more precisely, the cross-section software Drosg2000 [11] was employed. This software contains the designated data tables for several neutron energy ranges and neutron producing reactions.

The flux was determined twice: first, in front of the beaker glass at the target distance of the fission chamber and second, inside the sample cell for the subsequent activation calculations. After the MCNPX simulation, the fluxes were scaled to the experimentally determined flux of $\varphi=5.76\cdot10^6$ cm$^{-2}$s$^{-1}$ at the fission chamber. Following this, a CINDER90 simulation was employed to determine the induced activation in the sample. Since CINDER90 calculates the activation separately for each isotope, the results could easily be compared. The experimental and simulated results are presented in Table 1.

| Sample | Det. Isotope | Sim. Activ. $\frac{Bq}{kg}$ | Meas. Activ. $\frac{Bq}{kg}$ | Meas. Uncert. % | Deviation % |
|---|---|---|---|---|---|
| Ag | $^{108}Ag$ | 3.64E+05 | 2.51E+05 | 16.9 | 45.03 |
| Cu | $^{66}Cu$ | 2.48E+04 | 3.61E+04 | 3.1 | 31.30 |
| Cu | $^{60}Co^m$ | 1.67E+04 | 1.41E+04 | 14.9 | 17.90 |
| Cu | $^{64}Cu$ | 3.32E+03 | 4.91E+03 | 11.2 | 32.39 |
| Cu | $^{65}Ni$ | 9.64E+02 | 8.63E+02 | 3.2 | 11.75 |
| Cu | $^{62}Co$ | 1.45E+03 | 1.84E+03 | 44 | 21.30 |
| Al | $^{27}Mg$ | 4.62E+05 | 4.27E+05 | 5.4 | 26.60 |
| Al | $^{28}Al$ | 2.01E+04 | 2.74E+04 | 6.6 | 8.25 |
| Al | $^{24}Na$ | 5.14E+03 | 5.72E+03 | 2.5 | 10.27 |
| Fe | $^{56}Mn$ | 2.87E+04 | 1.51E+04 | 1.4 | 89.77 |
| Mn | $^{56}Mn$ | 6.09E+04 | 7.60E+04 | 1.4 | 19.89 |
| Mn | $^{52}V$ | 1.41E+04 | 1.16E+04 | 6.2 | 22.11 |

**Table 1.** Overview of the results of the activation experiment and the simulation

The first thing to note is a large discrepancy between the simulated and measured induced activation of iron, of around 90 %. After investigating the details of the CINDER90 procedure to calculate the isotope production, the reason became clear: The output of the MCNPX Cell Flux tally is, at the beginning of the CINDER90 calculations, rebinned to the 63 so called LaBauve energy group bins. For the production of the isotope $^{56}$Mn in the iron sample, this rebinning has a drastic consequence: Although the main flux is delivered at an energy of 8 MeV, this flux is grouped into a bin from 6.065 MeV to 10.000 MeV neutron energy for CINDER90 calculations. Most probably, also the cross-section database is binned in this fashion. Thus, if the cross-section for a certain reaction channel varies significantly within one energy window, the rebinning may cause a large deviation in the isotope production rate, either negative or positive, depending on whether the cross-section rises or falls with energy.

To check this postulate, the value of the cross-section for this reaction channel and bin was manually changed to the exact value at 8 MeV, taken from the EXFOR nuclear database, and the CINDER90 calculation was repeated. Thereby, the simulated value for the induced activation of the isotope $^{56}$Mn changed to $1.73 \cdot 10^4$ Bq/kg, which agrees much better with the measured value (the deviation is now 14 % instead of previously 90 %).

In conclusion, taking into account all possible sources for uncertainties, the benchmark experiment appears to validate the simulation, but one needs to be careful in interpreting the raw data, as the example of iron shows.

## 3. Simulations of Metal Samples in a Cotton Container

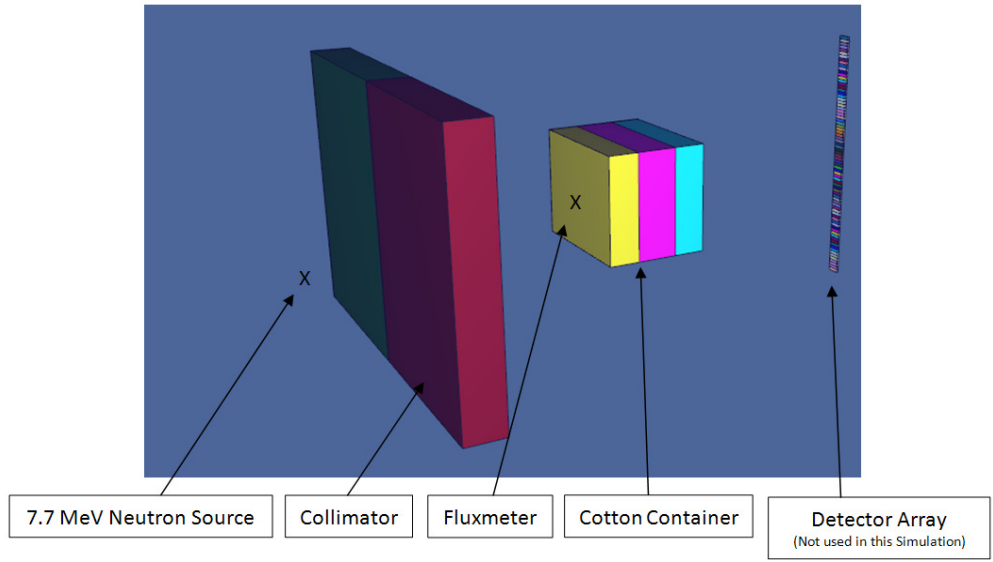

**Figure 3.** Schematic geometry of the MCNPX simulation

As air-freight goods shipped on passenger air planes are likely to contain appreciable quantities of either metallic or organic objects, the first simulation consisted in placing several metal samples (namely silver, gold, copper, aluminum and iron) in a cotton-filled container and irradiating it with neutrons. Because of technical reasons, the container was split into three parts. This simulation was repeated with an air-filled container, to determine the effect of neutron moderation and partial thermalization in the cotton. The scheme of the simulation setup can be found in Figure 3. A container of dimensions 201×153×164 cm (which are those of a standard LD-3 container neglecting the tapered bottom, which fits the contour of the cargo hold) was irradiated with a collimated fan beam (10×413.8 cm) at 400 cm distance from the center of the container to a monochromatic source of 7.7 MeV neutrons. Due to the large amount of cotton, the neutron energy spectrum broadens considerably by the time the neutrons impinge on the metal samples. The latter (see Figure 4), all around 200 g in mass, were placed at three axial positions: centered in the front, middle and back of the container.

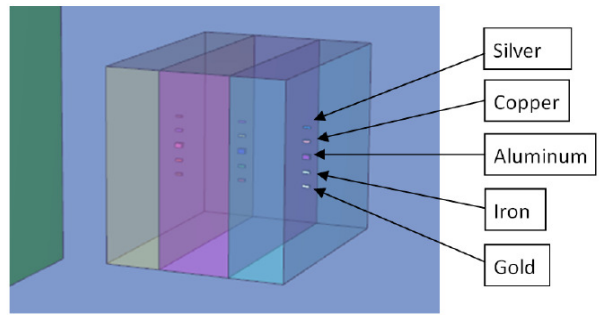

**Figure 4.** Metal samples in cotton container

As explained above, the simulations were performed with MCNPX combined with the activation code CINDER90. The latter makes use of the spectral flux distribution in the container and the samples, which is initially estimated by means of an MCNPX simulation. In a second step, the corresponding cargo geometry and material definition is invoked to calculate the isotopic inventory for each sample after the irradiation, and hence the resulting activities for each radioactive isotope species. To simulate an irradiation under realistic conditions, an irradiation time of 10 s was chosen. This corresponds to the proposed scanning time for a single frame, the full inspection of a regular air cargo container scanned by the ACCIS system being estimated at around 100 seconds. The required neutron fluence at the front face of the container for a single frame exposure over all energies is of the order of $10^6$ cm$^{-2}$.

Both, neutron induced activation of the samples as well as the ensuing gamma dose to water phantoms were simulated. The results will be presented in the following chapter.

### 3.1 Activation of Samples

The results of the activation simulation can be seen in Table 2 and Figure 5. Table 2 shows the specific activation of each sample at each position immediately (1 s) after the end of the irradiation, while Figure 5 shows the time dependence of the activation of the front samples thereafter.

| Material | Specific Activity $\left[\frac{Bq}{kg}\right]$ | | |
|---|---|---|---|
| | front Sample | mid Sample | back Sample |
| Silver | 1.97E+05 | 1.12E+05 | 2.63E+04 |
| Copper | 7.11E+02 | 5.81E+02 | 1.43E+02 |
| Aluminum | 2.67E+03 | 1.60E+03 | 3.76E+02 |
| Iron | 3.94E+01 | 5.32E+00 | 7.10E-01 |
| Gold | 7.64E+04 | 1.22E+04 | 1.86E+03 |
| Cotton | 1.37E-10 | 6.75E-11 | 1.61E-11 |

**Table 2.** Specific activation immediately after irradiation (cotton filled container)

The first thing to note is the activation of cotton. As a material containing oxygen, carbon and hydrogen, cotton is representative for most organic materials. The activation levels of cotton indicate that the neutron-induced activation of organic materials appears to be negligible. By comparison, the average human body contains around 4400 Bq, mainly from $^{40}$K, which corresponds to about 63 Bq/kg. On the other hand, the initial activity of the metallic samples is very high. However, most of the radionuclides produced decay very fast. The activation of the gold sample decreases by almost 4 orders of magnitude within the first 120 seconds; whereas that of the silver decreases by almost three orders of magnitude within 10 minutes.

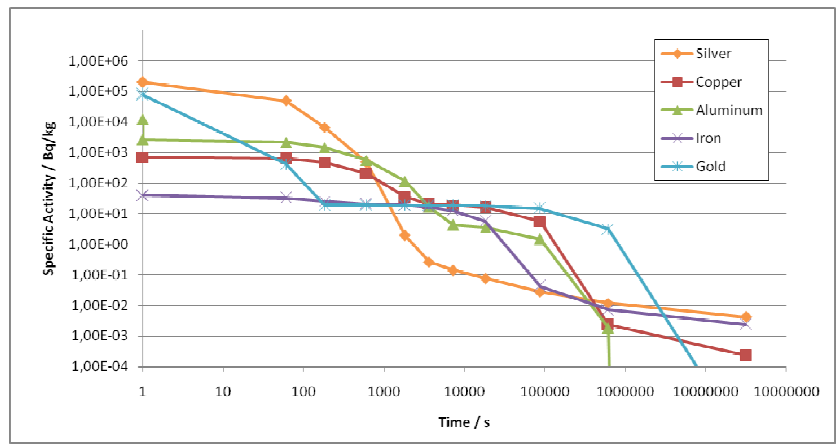

**Figure 5.** Time dependence of the activation of the front samples (cotton filled container)

The material which requires some more attention is aluminum. As every air cargo container consists of aluminum and containers are irradiated frequently during their lifespan, they could conceivably accumulate a significant radioactive inventory. Figure 5 shows that the specific activity of aluminum decreases strongly in the first second after irradiation (more than one order of magnitude). Afterwards, the activity is quite constant during about 20 minutes. However, after about one day the Al activation again decreases by around 3 orders of magnitude compared to its initial value. Consequently, repeated irradiation cycles will not lead to significant long-term accumulation of radioactive inventory in the container walls.

In the next step, the cotton inside the container was replaced by air, to avoid neutron moderation outside the metal samples. This scenario is, in a sense, at the opposite extreme, since the initial neutron spectrum is essentially unmoderated. The results of this simulation can be found in Table 3 and Figure 6.

| Material | Specific Activity $\left[\frac{Bq}{kg}\right]$ | | |
|---|---|---|---|
| | front Sample | mid Sample | back Sample |
| Silver | 4.62E+04 | 3.38E+04 | 2.57E+04 |
| Copper | 6.51E+01 | 4.62E+01 | 3.12E+01 |
| Aluminum | 1.55E+04 | 5.78E+03 | 2.24E+03 |
| Iron | 5.50E+01 | 3.90E+01 | 2.65E+01 |
| Gold | 9.63E+04 | 7.03E+04 | 5.32E+04 |

**Table 3.** Specific activity immediately after irradiation (air filled container)

In this case, two clear differences from the cotton-filled container are evident:
1. the induced activity in all three axial layers is more homogenously distributed. Specifically, in the cotton-filled container simulation, the activity of the back sample was approximately one order of magnitude lower than that of the front sample. However, in the simulation of the air-filled container, this axial-layer dependence is reduced to less than half an order of magnitude.

2. the overall activation levels for aluminum are significantly higher in the air-filled container than in the cotton-filled one. The same trend, but less pronounced, is observed for the iron and gold samples. In silver and copper, the trend is reversed.

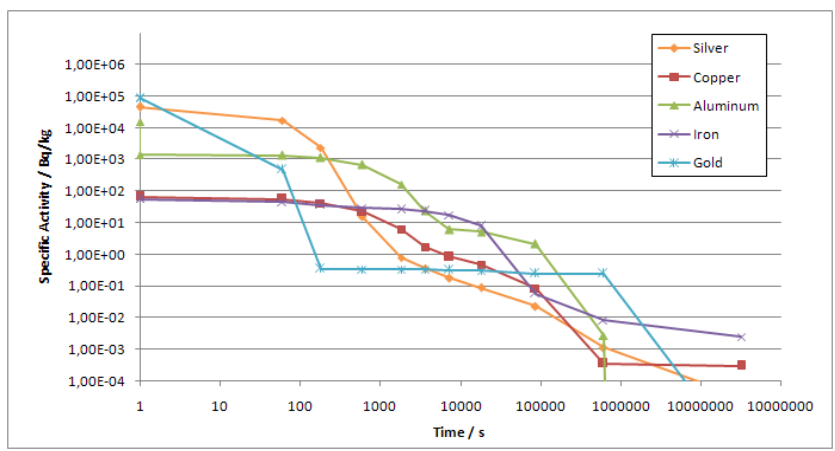

**Figure 6.** Time dependence of the activation of the front samples (air filled container)

This increased activation of the back-samples in the air-filled container is evidently due to the absence of moderation and screening by cotton. Thus, the induced activity of these samples is increased. Furthermore, the energy-behavior of the total elemental cross-sections is clearly reflected in the results. For example, compare the activity of the gold and copper samples with and without cotton: while the activity of gold stays approximately the same when comparing the empty with the cotton-filled container, the activity of copper decreases appreciably in the empty container. Comparing the total neutron cross-sections, the cross-section of gold is more or less flat (at several tens-of-barns) in the energy range down to a few eV, apart from several resonances in the 10 eV region. In contrast, the total cross-section of copper exhibits a large number of strong resonances in the keV region. This explains the higher activation level of copper for the cotton-filled container, in which the moderator effect reduces the neutron energy to values where the copper cross section is high.

**3.2 Induced Gamma Dose Rate**

To estimate the expected external dose rate by the induced activation in air cargo containers, we have simulated four scenarios relevant to specific groups of persons who may be exposed to radiation from inspected containers and (or) their contents following the irradiation. In the first scenario, the induced activation immediately after irradiation was taken into account for all 15 different samples. The dose recipients were 3 water phantoms (which serve as surrogates for human tissue) placed at three different positions (see Figure 7):

One phantom was placed in direct contact with the container (phantom 1), one at 1 m distance (phantom 2) and one at 3 m distance (phantom 3). The first scenario determines the maximal possible dose rate. The second scenario (1 m distance from container, 10 minutes after irradiation) represents a possible dose rate to cargo handlers, while the third scenario (3 m distance, 30 minutes after irradiation) simulates a situation relevant for the dose to passengers on the same aircraft as the inspected cargo, neglecting any available shielding. The fourth scenario (direct contact, 10 hours after irradiation) represents the possible exposure of a customer after the goods are delivered. The second, third and fourth scenarios were investigated only for the front samples and the phantom directly in contact with the container, since the result for all other setups can be estimated with the present data.

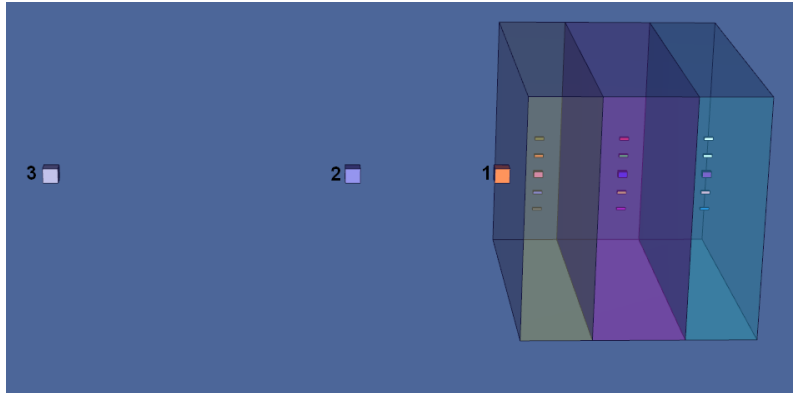

**Figure 7.** Overview of the phantom positions for the dose-rate simulation (all positions in one picture)

For the dose rate simulations, the CINDER90 script *gamma_source.pl* was employed to calculate an MCNPX gamma source definition for each activated metal sample and for the cotton contents of the front, middle and back part of the container. The sources are defined as homogeneous, isotropic gamma sources with element-dependent energy distributions within each metal sample cell. In an MCNPX calculation, these sources are employed to irradiate the three water phantoms of dimensions 10×10×10 cm$^3$.

In each simulation, the dose rate to the phantom in units of MeV · g$^{-1}$· s$^{-1}$ was measured, employing a tally measuring deposited energy in a specific cell. For ease of comparison, the results were converted to the unit μSv·h$^{-1}$·kg$^{-1}$, so the dose rate is normalized to the mass of the metallic samples in the container. Since only gamma radiation contributes, a weighting factor of 1 is chosen to convert energy dose to equivalent dose. The results of the first scenario can be found in Table 4.

| Sample | Dose rate to Phantom 1 $\left[\frac{\mu Sv}{h \cdot kg}\right]$ | | | Dose rate to Phantom 2 $\left[\frac{\mu Sv}{h \cdot kg}\right]$ | | |
|---|---|---|---|---|---|---|
| | front | middle | back | front | middle | back |
| Ag | 2.66E-03 | 1.84E-04 | 2.67E-05 | 1.42E-04 | 2.82E-05 | 6.35E-06 |
| Cu | 7.46E-05 | 3.21E-06 | 4.57E-07 | 2.73E-06 | 5.66E-07 | 1.19E-07 |
| Al | 1.32E-02 | 5.31E-04 | 7.96E-05 | 2.35E-04 | 4.04E-05 | 8.48E-06 |
| Fe | 3.79E-05 | 1.63E-06 | 2.35E-07 | 2.97E-07 | 6.06E-08 | 1.24E-08 |
| Au | 1.13E-03 | 8.43E-05 | 1.18E-05 | 1.39E-05 | 2.70E-06 | 5.91E-07 |

| Sample | Dose rate to Phantom 3 $\left[\frac{\mu Sv}{h \cdot kg}\right]$ | | |
|---|---|---|---|
| | front | middle | back |
| Ag | 5.02E-06 | 1.40E-06 | 3.97E-07 |
| Cu | 1.15E-07 | 3.55E-08 | 9.31E-09 |
| Al | 9.37E-06 | 2.63E-06 | 7.06E-07 |
| Fe | 7.07E-09 | 2.30E-09 | 5.79E-10 |
| Au | 2.42E-07 | 5.89E-08 | 1.43E-08 |

**Table 4.** Overview of dose calculations immediately after irradiation (cotton filled container). Please note the normalization of the dose rate which refers to 1 kg of sample mass !

To illustrate these results, a comparison can be made with the average equivalent dose per year received by a person in Germany (2.1 mSv/a), which corresponds to a dose rate of 0.24 µSv/h. An empty LD-3 container has a mass of around 72 kg and is made of aluminum. This would lead to an **initial** dose rate of 0.95 µSv/h which is 4 times the average annual exposure. Of course, this is just the dose rate immediately after the irradiation.

The time dependence of the dose rate is presented in Figure 8. As explained above, these simulations are performed just for the front sample, in which most of the radioactivity is induced and phantom 1, which is closest to this sample. In this example, the dose rate to sample 1 from the container would drop by three orders of magnitude within the first 10 hours after irradiation. The value for the dose rate after 10 hours is thus only 1.17 nSv/h.

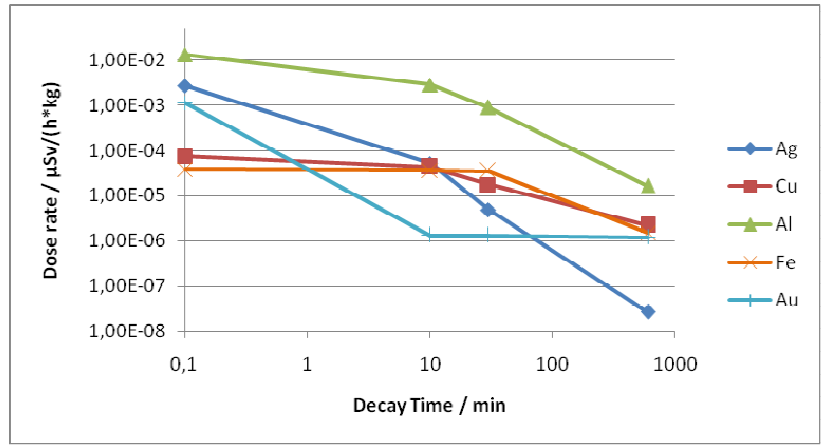

**Figure 8.** Time dependence of the dose rate per unit sample mass in phantom 1 of the front sample

In the next step, the dose rate arising from the cotton-filled container has to be compared to that of the empty container. For this purpose, the radiation transport simulation was repeated with the empty container, taking into account the front samples and phantom 1. A comparison of the results of an air-filled container to one fully loaded with cotton can be found in Table 5.

| Sample | Dose rate (Cotton) $\frac{\mu Sv}{h \cdot kg}$ | Activity (Cotton) $\frac{Bq}{kg}$ | Dose rate (empty) $\frac{\mu Sv}{h \cdot kg}$ | Activity (empty) $\frac{Bq}{kg}$ |
|---|---|---|---|---|
| Ag | 2.66E-03 | 2.02E+05 | 3.18E-05 | 4.62E+04 |
| Cu | 7.46E-05 | 7.13E+02 | 3.19E-05 | 6.51E+01 |
| Al | 1.32E-02 | 1.24E+04 | 1.31E-02 | 1.55E+04 |
| Fe | 3.79E-05 | 3.96E+01 | 5.85E-05 | 5.50E+01 |
| Au | 1.13E-03 | 8.35E+04 | 1.34E-03 | 9.63E+04 |

**Table 5.** Comparison of dose rates per unit of sample mass arising from the front side located metal samples in a cotton or empty container

It appears that there is no common trend for all samples in the comparison. For example, the dose rate due to the front silver sample is two orders of magnitude higher for the cotton filled container than for the empty container. On the other hand, the activity is less than one order of magnitude higher for the loaded container. The explanation is that the average gamma energy from activated silver in the cotton filled container is 237 keV while the one in the empty container is only 36 keV. The low energy gamma ray in the empty container is mainly reabsorbed by the silver sample itself and cannot contribute to dose outside the container. Comparing the dose rate of copper and silver in the empty container simulation, both values are almost equal, although the activation of silver is much larger. Also here the answer can be found in the gamma spectrum: Unlike the 36 keV average gamma energy, the gamma spectrum of activated copper sample has an average energy of 840 keV. In other words, the fraction of gamma rays which escape the sample and deposit energy in the phantom is much larger. An opposite effect occurs with iron or gold: The average gamma energies for the empty and cotton filled container are approximately equal at 1.12 MeV and 185 keV, respectively.

The cotton in the container has different effects on the external gamma field and the dose rate, which partially cancel each other: on one hand, the cotton moderates the neutrons. This can lead to a modified activation (which relates to the energy dependency of the materials neutron cross-sections). On the other hand, shielding of the gamma rays by cotton and the sample material itself needs to be taken into account. Naturally, this effect is sensitive to the gamma energy distribution which, in turn, is highly dependent on the materials and the energy spectra of the (partially) moderated neutron fields. In conclusion, a simulation with a simplified geometry is unlikely to yield results that are universally valid for an inspection facility of the kind described. Every single item within a container and its environment can modify the neutron spectrum (and thus, the gamma spectrum) and thereby also the dose rate outside the container. Therefore, the simulation presented can only yield a rough estimate of the expected dose rates.

## 4. Investigation of the Contribution of Neighboring Frames to the Activation

To perform a complete inspection of an object the size of an LD3 container, the full scan of the object will be divided into slices of around 10 cm width, so called frames. An effect neglected in previous simulations refers to activation arising from the irradiation of neighboring frames or, in other words, from scattered radiation to neighbor slices or frames

As movement of objects during the simulation is not incorporated in MCNPX or other recent Monte-Carlo toolkits, a workaround to simulate the contribution of inscattered radiation from neighbor frames to the total activation level had to be found. Although the object cannot be moved, the neutron source can. Thus, to simulate multiple fan beams, the collimator was equipped with two more slits and neutron sources in front of each slit were included. One source/slit system was shifted by 10 cm to the side (to simulate the first order neighbor frame), the other one by 20 cm (to simulate the next-nearest neighbor frame).

To prevent neutrons from one source traversing the slit in the collimator from another source, black (or fully absorbing, not scattering) collimators were placed in between the sources. With these multiple neutron sources, the same geometry as before (the cotton container including the metal samples) was irradiated. In each simulation, every neutron source was configured to start the same amount of source neutrons as in the previous simulation (that corresponds to a neutron fluence of $10^6$ cm$^{-2}$ in front of the container), to ensure comparability. Later on, also the third, fourth, fifth, sixth, seventh and eighth nearest neighbor frame was simulated by shifting the collimator slits and sources 30 cm, 40 cm, 50 cm, 60 cm, 70 cm and 80 cm off center. To compare the simulations, the activation fractions of the front aluminum sample were calculated employing CINDER90. As a result, the activation of the sample can be plotted over the source distance to the main frame, normalized to the activation due to the main source. This graph can be seen in Figure 9.

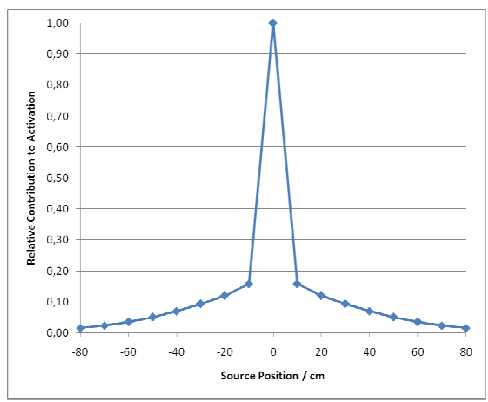

**Figure 9.** Contribution of neighbor frames to the activation level of a center frame

These results demonstrate that the total activation induced by the irradiation of neighbor frames amounts to about 115 % of the activation induced by the direct irradiation of the center frame. Consequently, this effect can by no means be neglected, although the simulated geometry is an extreme case in terms of neutron scattering, due to the cotton filling in the container.

## 5. Gamma Induced Activation

As explained in the introduction, the main goal of the ACCIS project is the combination of two nuclear interrogation techniques, namely Fast-Neutron Resonance Radiography (FNRR) and Dual Discrete-Energy γ-Radiography (DDEGR), in one interrogation system. In other words, two kinds of radiation, neutrons and gammas, will be employed.

In this work, the main focus was on the simulation of neutron induced activation and dose. However, energetic gamma-rays can also induce radioactivity via photonuclear reactions. Although the thresholds for these reactions are at rather high gamma energies (≳ 5 MeV), 15.1 MeV gamma rays arising from the $^{11}B(d,n+\gamma)^{12}C$ reaction are capable of inducing these reactions. Therefore, also the gamma induced activation needs to be estimated.

The main problem of simulating gamma induced activation is the lack of simulation tools with suitable cross-section data. In MCNPX no nuclei are produced or transported. CINDER90 calculates only neutron-induced activation. In Geant4, calculated cross-sections of photo-nuclear reactions for about 50 nuclei are included. The data are based on analytical models taking into account the predominant Giant Dipole Resonance cross sections; see [12] or [13]). However, only nuclide production (but no dose tally) is available in Geant4. Furthermore, also in evaluated cross-section databases, such as EXFOR[14], data tables for many inelastic reaction channels in the relevant energy range are missing.

Within the scope of this paper, only a schematic outline of how these issues can be addressed will be presented. Even so, the preliminary results obtained conclusively demonstrate that gamma-induced activation is almost negligible compared to neutron-induced activation. Two approaches to simulate the gamma induced activation were elaborated during this work. Both will be explained in the following, employing an aluminum sample inside the cotton container, similar to the previous simulations with neutrons. However, this time the sample and the container are irradiated with 15 MeV gamma rays.

In the $^{11}B(d,n+\gamma)^{12}C$ reaction the gamma flux was determined to be about one order of magnitude smaller than the neutron flux.. Furthermore, with deuteron beams of 5 – 7 MeV, the thick-target yield of the 4.43 MeV gamma rays is approximately equal to that of the 15.11 MeV [Brandis et al, to be published]. This is also in accordance with the requirements for the proposed cargo inspection technique. The expected gamma flux for an investigation is therefore around $5 \cdot 10^4$ cm$^{-2}$ s$^{-1}$.

For this investigation, the aluminum sample was chosen for two reasons: On one hand, aluminum turned out to be the most critical material concerning neutron activation in the previous simulations. On the other hand, aluminum is a rather simple case for gamma activation, as it is mono-isotopic and gives rise to just three photo-nuclear reaction channels $^{27}$Al($\gamma$,n)$^{26}$Al, $^{27}$Al($\gamma$,p)$^{26}$Mg and $^{27}$Al($\gamma$,$\alpha$)$^{23}$Na. Of these, only one resulting nucleus is radioactive, namely $^{26}$Al, which is long lived $T_{1/2}=7.17 \cdot 10^5$a.

First we simulated the gamma induced activity with MCNPX and CINDER90: In the initial step the gamma flux through the aluminum sample is simulated with a regular MCNPX simulation, employing 15 MeV gamma rays. The problem faced in the next step with CINDER90 is that only neutron cross-sections are included and consequently it allows only neutrons as input particles. To bypass these constraints, the cross-section tables were supplemented "by hand" with photo-nuclear cross-section data taken from the EXFOR database [14] and furthermore, the MCNPX output file was camouflaged to present a CINDER conform neutron simulation. Having done that, a conventional CINDER90 calculation can be started. For the given example of the aluminum sample in the cotton container, 305 $^{26}$Al nuclei were produced during the irradiation of one frame, which leads to an initial activation of $3.58 \cdot 10^{-11}$ Bq/kg after a 10 sec. irradiation. In principle, CINDER90 could now be employed to produce a secondary gamma source definition for further dose calculations by MCNPX, but as the induced activation turns out to be quite insignificant compared to the neutron activation (which resulted in an initial activity of $2.67 \cdot 10^3$ Bq/kg), this step was skipped. The more fundamental problem encountered when employing this method for gamma rays is the lack of available cross-section data. For irradiation of aluminum, only ($\gamma$,p) and ($\gamma$,n) cross-section tables were available.

The second approach to perform gamma-induced activation calculations employs Geant4, CINDER90 and MCNPX: A Geant4 simulation is prepared, with exactly the same geometry as in the previous neutron-based MCNPX simulation. For this simulation it is very important that the so called "G4PhotoNuclearProcess" is included in the physics list. Geant4 can now be employed, to calculate the isotope density of the daughter nuclides generated during the gamma irradiation. The initial isotope densities, calculated by Geant4 are translated into an MCNPX conform output file (equivalent to the above mentioned MCNPX output file of a comparable neutron simulation). This output file is then used in CINDER90 to calculate the initial induced activity as well as the decay of the isotopes calculated by the initial Geant4 calculation.

Employing this method to estimate the gamma induced activation of the aluminum sample 208 $^{26}$Al nuclei were obtained, which leads to an initial activation (after a 10 sec.-irradiation) of $2.45 \cdot 10^{-11}$ Bq/kg.

Obviously, the two methods lead to results which differ by around 46 %. This can be explained by the fact that, in the first approach, a comprehensively evaluated cross-section was employed (EXFOR), while the cross-section in the second was semi-empirical in nature and only took into account part of the relevant physics [13].

From the practical point of view the radio-isotope inventory and activation levels in gamma induced activation are low compared to neutron-induced activation (in the case of aluminum about $10^{14}$ times lower). Therefore the focus on neutrons in this work is justified.

## 6. Summary

The objective of the present work was to simulate neutron induced activation in cargo by a neutron- and gamma- based radiographic air freight inspection system. To achieve this goal, a combination of the particle transport code MCNPX and the transmutation code CINDER90 was employed. In this combination, MCNPX simulates the neutron flux through a certain cell and subsequently, CINDER90 uses the results, as well as the geometry definition, to calculate the isotope inventory and thus the activation. Furthermore, CINDER90 is able to create an MCNPX gamma source definition to perform the dose simulations. Using these computational tools, the radiation dose emanating from activated cargo was estimated.

The computer simulation model was verified by a benchmark experiment using samples of iron (Fe), copper (Cu), silver (Ag), aluminum (Al) and manganese (Mn) and irradiating them by fast neutrons. The experimental results are in reasonably good agreement with the principal features of the simulated values, however, the crude binning of the neutron flux and the cross-section tables in CINDER90 occasionally lead to significant deviations. Thus, care and attention are required when interpreting the results of the simulation. Evidently, this is an issue which should be addressed in future developments of the CINDER program package.

In the next step, simulations under realistic irradiation assumptions for an air cargo inspection facility were performed. They consisted in neutron-activation of metal samples (aluminum, copper, silver, gold and iron) enveloped in a cotton- or air-filled container. Immediately after irradiation, the samples showed initial activations in the range between $(4.0 \cdot 10^1 - 2.0 \cdot 10^5)$ Bq/kg sample mass, depending on the sample material. Furthermore, it could be shown that the activity decays by around 2 orders of magnitude within the first 10 minutes. These values need to be compared to "natural" radioactivity levels, e.g. of concrete ($3.9 \cdot 10^2$ Bq/kg $^{40}$K) or a human body ($4.4 \cdot 10^3$ Bq/kg $^{40}$K), which are of the same order of magnitude.

Finally also the gamma dose rates from the activated samples were determined for different scenarios, encompassing various groups of people who may be exposed to radiation from the container and (or) its contents at different times after the inspection. It was shown that the dose rates of all samples immediately after the irradiation is below 0.02 µSv/h per kg sample mass for the cotton-filled container. For example, the interrogation of an empty LD-3 container, made of aluminum and having a mass of 72 kg, would lead to an initial dose rate of about 0.95µSv/h, which is about a factor of 4 higher than the average effective natural dose rate received by a person in Germany. However, the neutron-induced dose rate decreases by about 2 orders of magnitude within 30 minutes after the irradiation is completed. Thus it is only relevant for personnel which have direct physical contact in this phase of the logistic process.

It was also shown that the differences in simulated dose rates between a cotton-filled and an air-filled container strongly depends on the gamma spectrum of the activated samples, since the cotton acts as a neutron moderator as well as a shield against gamma-induced radiation.

In a third simulation the effect of activation due to scattered neutrons from neighboring frames was calculated. From this simulation it was concluded that this is a significant contribution and adds up to 115% of the primary, direct activation. Finally, as the proposed ACCIS system is a combined neutron- and gamma-based interrogation system, the outline of a method for investigating the effect of gamma-induced activation was presented. Since the CINDER90 code is only able to calculate neutron induced activation different simulation techniques were required. As an example, the gamma-induced activation of an aluminum sample, which is wholly associated with the isotope $^{26}$Al, was calculated using two different methods. The result is an activation of about $3.02 \cdot 10^{-11}$ Bq/kg. As this activation is several orders of magnitude lower than the neutron activation, it was concluded that photonuclear activation is not relevant in this scenario and thus, further dose calculations are not required.

## Acknowledgement

This work is partially performed within the research project "ACCIS", which is funded by the German Ministry of Education and Research (BMBF) and the Israeli Ministry of Science and Technology (MOST) in the joint program "Research for Civil Security". Furthermore, this work is supported by NATO within the "Science for Peace" project SFP-983150